\documentclass{elsart}

\usepackage{graphicx}
\usepackage{amssymb}
\usepackage{epsfig}

\def\Journal#1#2#3#4{{#1} {#2} (#3) #4}

\def\NPA{Nucl. Phys. {\bf A}}
\def\PLB{Phys. Lett.  {\bf B}}
\def\PRL{Phys. Rev. Lett.}
\def\PRC{Phys. Rev. {\bf C}}
\def\PRD{Phys. Rev. {\bf D}}

\def\JPG{J. Phys. {\bf G}}

\def\be{\begin{equation}}
\def\ee{\end{equation}}

\newcommand{\ud}{\mathrm{d}}

\begin{document}
\begin{frontmatter}

\title{Charmonium from Statistical Hadronization of Heavy Quarks -- a Probe
  for Deconfinement in the Quark-Gluon Plasma}

\author[gsi,tud]{P.~Braun-Munzinger},
\author[hei]{J.~Stachel}

\address[gsi]{ExtreMe Matter Institute EMMI, GSI Helmholtz Zentrum f\"ur
  Schwerionenforschung,   
D-64291 Darmstadt, Germany}
\address[tud]{Technical University Darmstadt, D-64289 Darmstadt, Germany}
\address[hei]{Physikalisches Institut der Universit\"at Heidelberg,
D-69120 Heidelberg, Germany}

\begin{abstract} 
  We review the statistical hadronization picture for charmonium production in
  ultra-relativistic nuclear collisions. Our starting point is a brief
  reminder of the status of the thermal model description of hadron production
  at high energy. Within this framework an excellent account is achieved of
  all data for hadrons built of (u,d,s) valence quarks using temperature,
  baryo-chemical potential and volume as thermal para\-me\-ters. The large
  charm quark mass brings in a new (non-thermal) scale which is explicitely
  taken into account by fixing the total number of charm quarks produced in
  the collision.  Emphasis is placed on the description of the physical basis
  for the resulting statistical hadronization model. We discuss the evidence
  for statistical hadronization of charmonia by analysis of recent data from
  the SPS and RHIC accelerators.  Furthermore we discuss an extension of this
  model towards lower beam energies and develop arguments about the prospects to observe
  medium modifications of open and hidden charm hadrons. With the
  imminent start of the LHC accelerator at CERN, exciting prospects for
  charmonium production studies at the very high energy frontier come into
  reach. We present arguments that, at such energies, charmonium production
  becomes a fingerprint of deconfinement: even if no charmonia survive in the
  quark-gluon plasma, statistical hadronization at the QCD phase boundary of
  the many tens of charm quarks expected in a single central Pb-Pb collision
  could lead to an enhanced, rather than suppressed production probability when
  compared to results for nucleon-nucleon reactions scaled by the number of
  hard collisions in the Pb-Pb system.
\end{abstract}

\vspace{2mm}

\end{frontmatter}

\section{Introduction}

Investigation of hadron production in ultra-relativistic
nucleus-nucleus collisions has revealed convincing evidence for a
thermal production mechanism.  In particular, the study of yields of
hadrons composed of light (u,d,s) valence quarks from AGS up to RHIC energies
has shown
\cite{agssps,satz_had,heppe,cley,beca1,rhic,nu,beca2,rapp,becgaz,aa05,man08}
that hadron multiplicities can be described quantitatively in the
framework of a hadro-chemical equilibrium approach. Within this model
the only parameters are thermal quantities: the chemical freeze-out
temperature $T$, the chemical potentials $\mu$ and, if
applicable, the fireball volume $V$; for a recent review see
\cite{review}.

The underlying picture of the evolution of the system formed in a
collision between two heavy nuclei at high energies is the following:
In the early phase of the collision partons are liberated in hard
collisions describable by perturbative quantum chromo dynamics
(QCD). The partonic system subsequently equilibrates, i.e. reaches
(approximate) local momentum isotropy, all the while expanding in beam
direction with velocity of light. This system is called the fireball
and is characterized by thermal parameters such as a temperature and
by an equation of state. Eventually also transverse expansion builds
up. The expanding fireball cools with its temperature dropping as T
$\propto \tau^{-1/3}$ (or slightly faster due to transverse
expansion). Eventually the fireball reaches the phase boundary between
quarkmatter and hadronic matter and the partonic degrees of freedom
are converted into hadronic degrees of freedom. The corresponding
reduction in degrees of freedom is more than a factor of 3 and
therefore the volume has to grow accordingly during hadronization. At
some temperature equal or below the critical temperature hadron yields
are frozen in. This is what is called the chemical freeze-out and the
corresponding thermal parameters T and $\mu$ are determined from the
analysis of hadron yields as discussed in the previous paragraph \footnote{We
  have no direct evidence that the quark-gluon plasma before hadronization is
  in chemical equilibrium although this is likely the case for the light quark
flavors and maybe also between light quarks and gluons.}. The
now hadronic fireball may expand and cool further until elastic
collisions seize to change the momentum distributions. This point is
called thermal freeze-out. After this point there may still be some
residual interactions (e.g. Coulomb interaction) and weak decays. The
resulting momenta and particle types are measured in the detectors.
 
An important outcome of the hadron yield investigations is that the
extracted temperature values rise sharply with energy for energies in
the c.m. system per colliding nucleon pair $\sqrt{s_{NN}}$ below 10
GeV and reach constant values near $T$=160 MeV; this appears to
be the limiting temperature for a hadronic fireball. The
baryo-chemical potential exhibits a strong (monotonous) decrease as a
function of energy. Even detailed features such as the, up to very
recently considered rather mysterious, sharp maximum in the excitation
function of the $K^+/\pi^+$ ratio at SPS energies (``the horn'') can
be described quantitatively if full account is taken of the hadronic
mass spectrum \cite{horn}. The limiting temperature implies a
connection to the QCD phase boundary and it was, indeed, argued
\cite{wetterich} that the quark-hadron phase transition drives the
equilibration dynamically, at least for SPS energies and above.

The success in describing the production of hadrons containing light
quarks in a thermal framework raises the question whether hadrons
containing charm (or even bottom) quarks could also be produced
thermally. The answer at first glance is no: the charm quarm mass $m_c
\approx 1.3$ GeV brings in a new scale which exceeds significantly any
conceivable temperature which may be reached in nucleus-nucleus
collisions. It also is much larger than $\Lambda_{QCD}$, implying that
charm production should be describable in a perturbative QCD
framework. Thermal production of charm quarks is thus at best (at LHC
energy) a small perturbation compared to production in hard
scattering, and at presently available (RHIC and SPS) energies
completely negligible. It is thus not surprizing that hadrons
containing charm quarks, and in particular charmonia were considered
key probes to diagnose the fireballs formed in ultra-relativistic
nucleus-nucleus collisions.

Charmonia are bound states of charm quarks and their antiquarks. In
nuclear collisions at high energy,  charm quarks can be produced
rather copiously, leading in turn to significant cross sections for
charmonium production. The typical radii of charmonium mesons are of
order 0.2 - 0.5 fm, i.e. comparable to or larger than the mean
distance between partons in a Quark-Gluon Plasma (QGP) of temperature
of a few hundred MeV. Because of this fact, charmonium production is
considered an important probe to determine the degree of
deconfinement reached in the fireball produced in ultra-relativistic
nucleus-nucleus collisions. In fact, Matsui and Satz argued in
\cite{satz} that the expected high density of gluons in a QGP formed
by collisions between heavy nuclei at ultra-relativistic energy should
destroy any charmonia formed prior to the QGP in a process analoguous
to Debye screening in an electromagnetic plasma. Consequently,
suppression of charmonia (compared to their production in the absence
of QGP) was proposed as a ``smoking gun'' signature for plasma
formation in nuclear collisions at high energy. Measurements at the
CERN SPS accelerator indeed provided evidence for such suppression
\cite{na50} in central collisions between heavy nuclei, while no
suppression was found in more grazing collisions or collisions between
light nuclei, where plasma formation is not
expected. However, absorption of charmonium in the nuclear medium as
well as its break-up by hadrons produced in the collision were also
identified as possible mechanisms leading to charmonium suppression
even in the absence of QGP formation
\cite{gavin,capella,wong,urqmd_jpsi}, and the interpretation of the
SPS data remains inconclusive. Recent measurements performed at the RHIC
accelerator yielded a $J/\psi$ suppression similar to that observed at the
SPS, leading to further puzzles. For a recent review see \cite{raphael_gr}.

Furthermore, it was noted in \cite{gaz-gor} that charmonium and, in
particular, J/$\psi$ production in Pb-Pb collisions at SPS energy
exhibit thermal features. Considering these and the general success of
the thermal model in describing hadron production, but keeping in mind
the new scale brought into the problem by the charm quark mass, led us
to propose, in 2000, the statistical hadronization model
\cite{pbm1}. In this approach, the number of charm quarks plus their
antiparticles is effectively decoupled from the thermal description of
charm quark hadronization at the QCD phase boundary, where hadrons
are formed from charm quarks with statistical weights calculated at
chemical freeze-out. In \cite{pbm1} we have assumed that all charm quarks
are formed in initial hard collisions but other processes such as
thermal production in extremely hot fireballs formed at LHC energy
\cite{ko-lhc} could be considered in addition. Apart from the total
number of charm quarks, the only parameters entering the statistical
hadronization model are the thermal parameters which are very well
constrained by chemical freeze-out analyses.

The idea of statistical hadronization led to a series of
investigations of $J/\psi$ production based on this approach
\cite{gor1,gra1,kos,aa1,bra,aa2,aa3}.  Initial interest focussed on
the available SPS data for $J/\psi$ production in Pb-Pb collisions,
but the trends for RHIC and LHC were also investigated \cite{aa2,aa3,rapp08}.
We note that a different approach, based on a kinetic model
\cite{the1,the2,the3} has been developed. Here, charmonia are
continously formed (and destroyed) during the plasma phase and
dynamical input, such as J/$\psi$ production cross sections, is
needed to make predictions. Recently the statistical hadronization
model was extended to the production of $\Upsilon$ mesons
\cite{gra2,aa2} and multiply heavy-flavored hadrons \cite{bec_hq}.
Very comprehensive reviews of quarkonium production in heavy
ion collisions can be found in \cite{rapp_cr,bhaduri}.

A key feature of the statistical hadronization model as well as of the
kinetic model is the fact that yields for the production of charmonia
scale, because the charmonia are combined from charm and anti-charm
quarks, with the square of the number of charm quarks $N_c$ in the
system. This implies that, for sufficiently large $N_c$, the
production of charmonia should be enhanced \cite{pbm1,aa1,aa2,aa3}
rather than suppressed in a fully deconfined QGP.

An important aspect of charmonium production and its possible
modification in nuclear collisions is the sequence of time steps and
scales involved in the process. In the original proposal \cite{satz}
charmonium (or, at least, its precursors, such as a color-octet c$\bar
c$ state) is produced early, before quark-gluon plasma formation and
also in a time interval which is short compared to the collision
time. Conversely, in the statistical hadronization model, charmonia
and all charmed hadrons are formed late, at the time of hadronization
of the QGP.  In view of these wildely different approaches a thorough
review of the relevant time scales was performed recently
\cite{charm_le} and will be important input for the understanding of
charmonium as a probe for deconfinement in ultra-relativistic
nucleus-nucleus collision.

We start this review with a brief reminder, in section 2, of the
current status of the thermal model for hadron production. An account
of the statistical hadronization model for charmonia and charmed
hadrons is given in section 3.  In this section  we will also  discuss
the various time scales relevant for charm, charmonium, and open charm
hadron production and discuss their relevance for the applicability of
the statistical hadronization model as well as for the study of
possible medium effects in the charm sector.  Section 4 will deal with
the current status of charmonium production in nucleus-nucleus
collisions at SPS and RHIC energies and its interpretation in terms of
the statistical hadronization model. Our results on open charm and
charmonium production and their sensitivities to possible in-medium
modifications, in particular at energies below top SPS energy, will be
presented in section 5. Section 6 will be devoted to predictions for
LHC energy.

\section{Status of thermal production of hadrons in relativistic
  nucleus-nucleus collisions}

Hadron production and in particular hadron yields have been studied
with high precision in central nuclear collisions at AGS, SPS, and
RHIC energies.  These yields can be described by assuming that all
hadrons are formed when the fireball reaches a certain temperature T,
chemical potential $\mu_b$\footnote{Other chemical potentials are fixed by
  conservation laws and are thus no free parameters.}, and volume. Under these conditions the
yields can simply be described by the thermodynamical grand canonical
(or in special cases canonical) ensemble. The underlying process is
called ``chemical equilibration'' in analogy to particle production in
the early universe.  Recent detailed analyses have been performed by
\cite{becgaz,aa05,horn}, a comprehensive review can be found in
\cite{review}. Importantly, the data at SPS and RHIC energies comprise
yields of multi-strange hadrons including the $\Omega$ and $\bar
\Omega$ baryons. Their yields agree very well with the chemical
equilibrium calculation and are strongly enhanced as compared to
observations in pp collisions. This implies that, at chemical freeze-out, the
strangeness sector is in equilibrium as well.

\begin{figure}[hbt]
\begin{tabular}{lr} \begin{minipage}{.49\textwidth}
\centering\includegraphics[width=1.\textwidth]{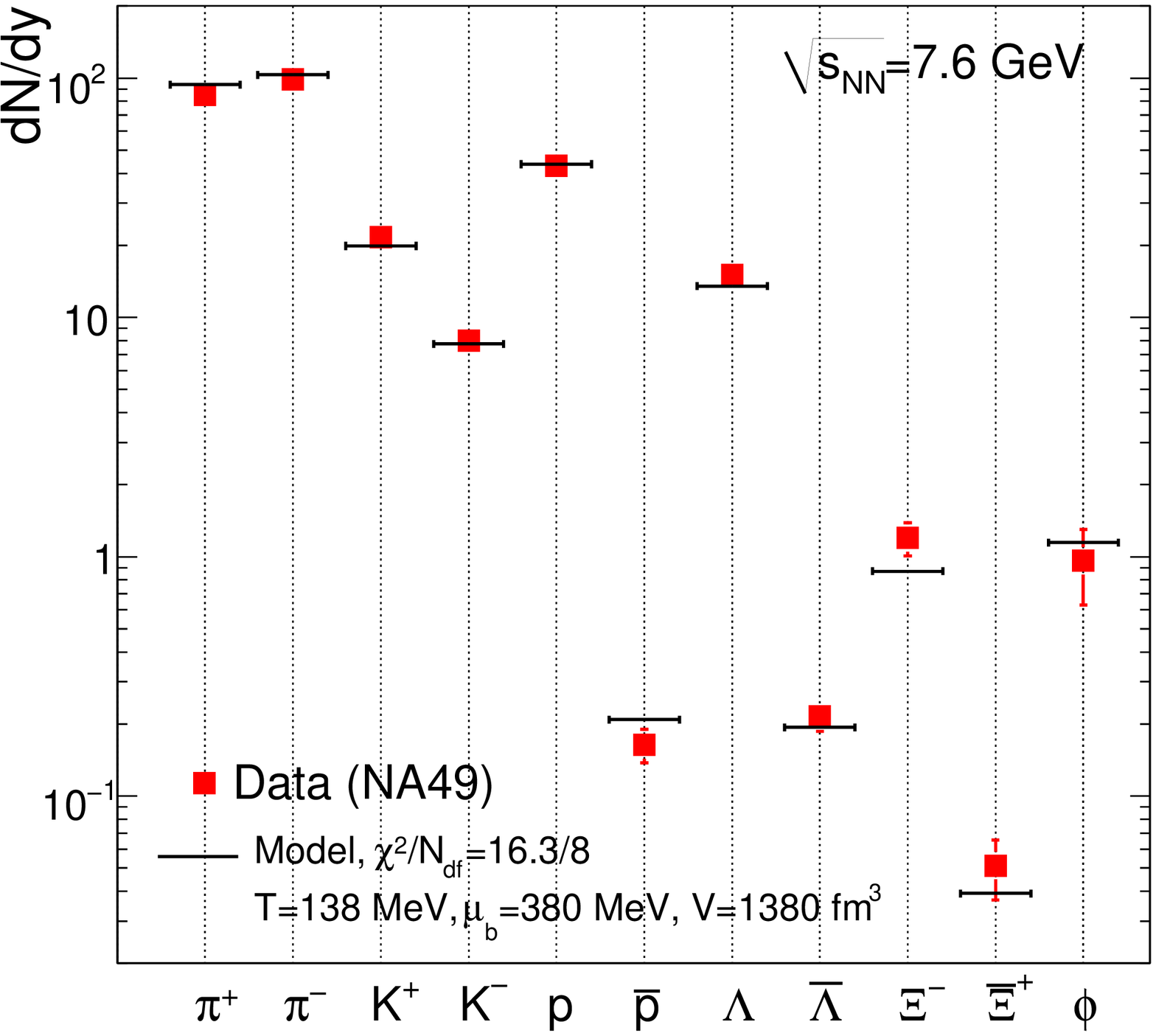}
\end{minipage} &\begin{minipage}{.49\textwidth}
\centering\includegraphics[width=1.\textwidth]{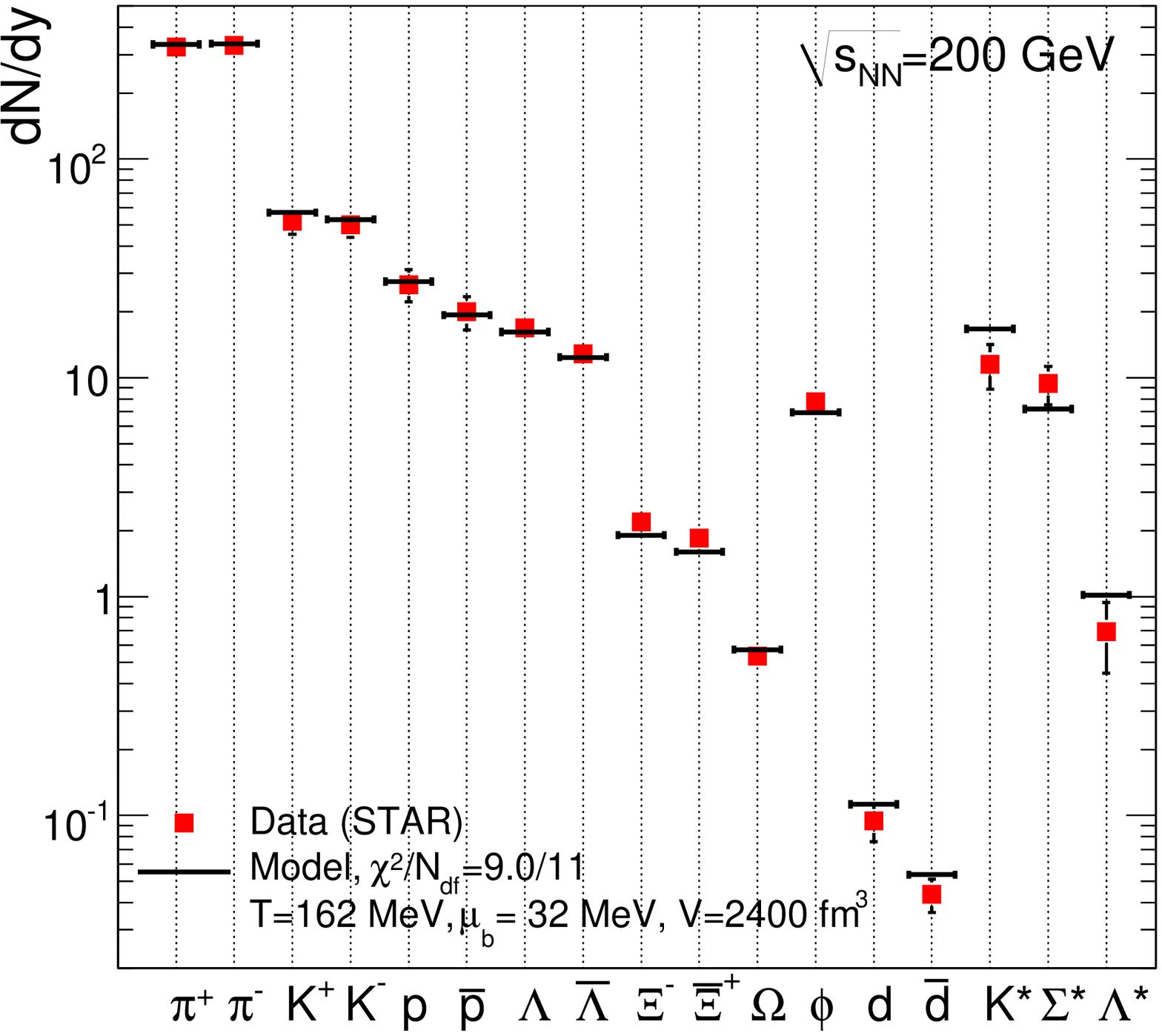}
\end{minipage} 
\end{tabular}
\caption{Experimental hadron yields and model calculations for the
  thermal parameters of the best fit at the c.m. energies per
  colliding nucleon pair of 7.6 (left panel) and 200 GeV (right
  panel). The strongly decaying resonances were not included in the fit.}
\label{fig_fits}
\end{figure}

In Fig.~\ref{fig_fits} we present a comparison of measured and
calculated hadron yields at the energies of $\sqrt{s_{NN}}$=7.6 GeV
(beam energy of 30 A GeV at the SPS) and $\sqrt{s_{NN}}$=200 GeV. The
results are taken from the most recent analysis of \cite{horn}. The
data sets are very well reproduced by the model calculations and this
applies to all energies, from the lowest AGS beam momentum of 2 A GeV
up to the top RHIC energy of $\sqrt{s_{NN}}$=200 GeV.

\begin{figure}[hbt]
\centering\includegraphics[width=.75\textwidth]{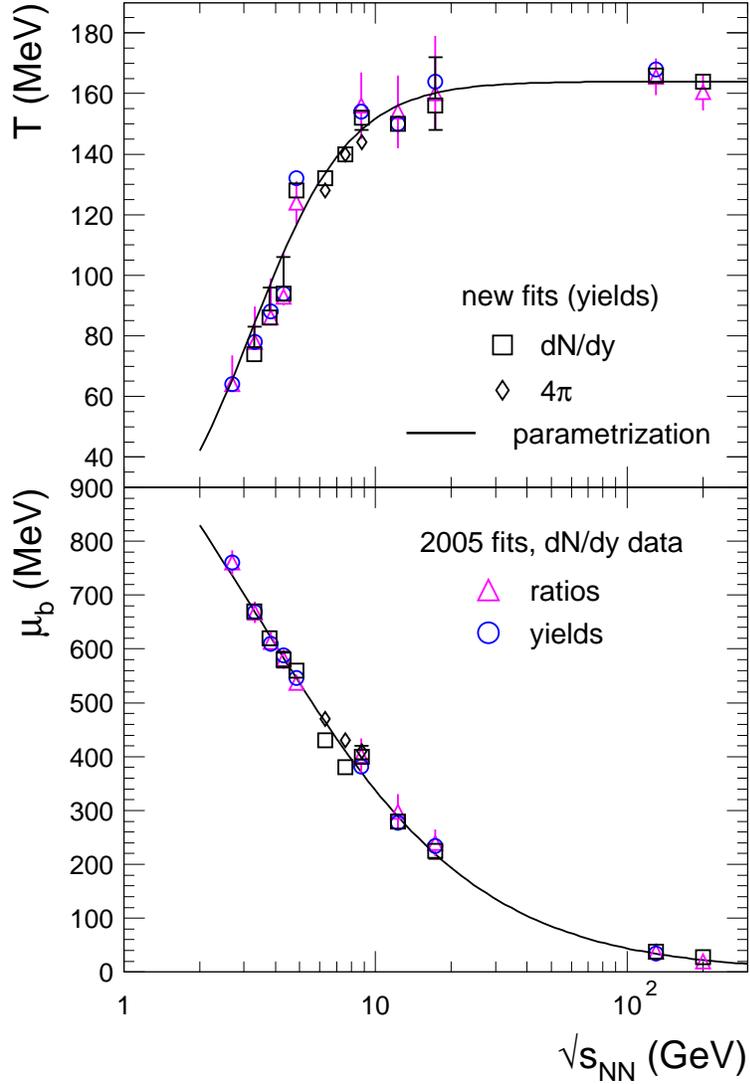}
\caption{The energy dependence of temperature and baryon chemical
  potential for hadron formation. The most recent results (open squares and
  diamonds) are compared
  to the values obtained in an earlier study \cite{aa05} (open triangles and
  circles). The lines 
  are parametrizations of the most recent results for $T$ and
  $\mu_b$. Figure taken from \cite{horn}.}
\label{fig_tmu}
\end{figure}

The energy dependence of the so-obtained ``chemical'' parameters T and
$\mu_b$ is shown in Fig.~\ref{fig_tmu}. In contrast to the smoothly
decreasing chemical potential the temperature parameter first
increases strongly with energy but saturates rather abruptly near an
energy of $\sqrt{s_{NN}}$ = 10 GeV at a value slightly above 160
MeV. This behavior strongly supports Hagedorn's limiting temperature
hypothesis \cite{hagedorn} and lends support to the notion that the
saturation behavior is caused by the QCD phase boundary which,
according to lattice QCD calculations, runs for $\mu_b \leq $ 300 MeV
at an approximately constant temperature.

The impressive agreement shown in Fig.~\ref{fig_fits}
extends also to rather delicate features of hadron production such as
the energy dependence of the $K^+/\pi^+$ and $\Lambda/\pi^+$ ratios as
demonstrated in \cite{horn}. For the purpose of the present review we
note that the thermal parameters $T$ and $\mu_b$ as well as the volume
which serve as input into statistical hadronization model
calculations are completely determined at all energies\footnote{For LHC
  energy we assume $T= 164$ MeV and $\mu_b= 1$ MeV in obvious
  extrapolation of the SPS and RHIC data.}. Parametrizations for both are
given in \cite{horn}.

\begin{figure}[htb]
\begin{tabular}{lr} \begin{minipage}{.6\textwidth}
\centering\includegraphics[width=1.15\textwidth]{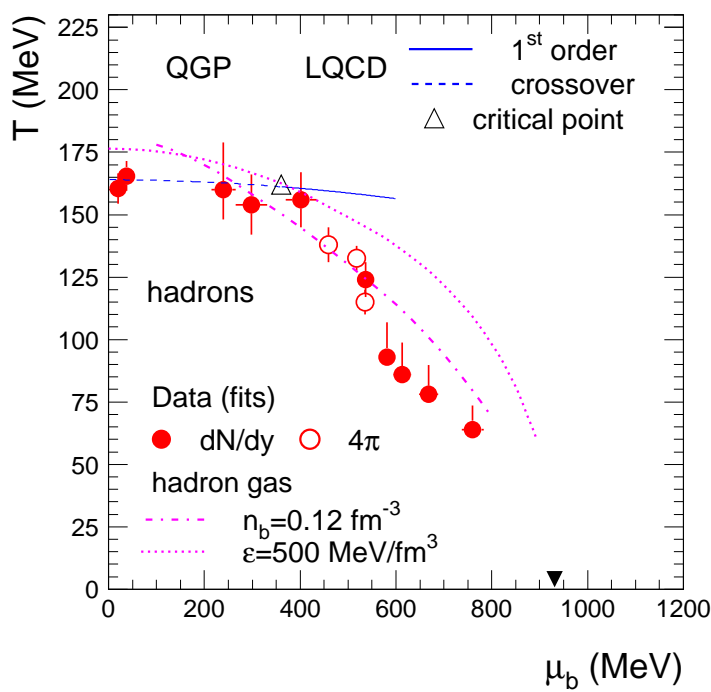}
\end{minipage} &\begin{minipage}{.37\textwidth}
\caption{The phase diagram of nuclear matter in the $T$-$\mu_b$ plane.
  The experimental values for the chemical freeze-out are shown
  together with results from lattice QCD calculations \cite{lqcd}.
  The critical point found in the LQCD calculation is marked by the
  open triangle.  Also included are calculations of freeze-out curves
  for a hadron gas at constant energy density ($\varepsilon$=500
  MeV/fm$^3$) and at constant total baryon density
  ($n_b$=0.12~fm$^{-3}$). The full triangle indicates the location of
  ground state nuclear matter (atomic nuclei). Figure taken from \cite{aa05}.}
\label{fig17}
\end{minipage} \end{tabular}
\end{figure}

To emphasize the connection to the QCD phase transition we present the
values of $T$ and $\mu_b$ obtained from our fits of the experimental
data in a phase diagram of hadronic matter in Fig.~\ref{fig17}.  An
important observation about the phase diagram is that, for the 40 A
GeV SPS energy and above, the thermal parameters for hadron formation
agree with the phase boundary calculations from lattice QCD (LQCD)
\cite{lqcd,lqcd1,lqcd2}, implying that hadron yields are frozen near
the phase boundary.

The LQCD calculation \cite{lqcd} shown in Fig.~\ref{fig17} is for two
light quarks ($u, d$) with realistic (close to physical) masses and a
heavy strange quark.  The critical temperature at $\mu_b$=0 from LQCD
calculations is about 170 MeV \cite{fk,fodor07} (and refs. therein),
with a scale uncertainty of the order of 10 MeV \cite{fk,lqcd2} but
with significany systematic errors \cite{fk,fodor07}.

Despite significant advances in high performance computing many aspects of the
QCD phase transition remain unclear. While the authors of \cite{fodor_nature} argue that the
transition is of cross-over type, implying the existence of a critical point,
this is disputed on very general grounds in \cite{digiacomo} (see also
\cite{pbmjw} for a general survey). Furthermore,  precise
values for the transition temperature do not exist today. In fact,
results ranging from 150 to 190 MeV have been obtained for $T_c$. The
situation is well summarized in \cite{fodor07,cheng06}; according to
these authors the range in temperatures could be due to a cross-over
nature of the phase transition for small values of $\mu_b$
\cite{fodor07}. Here, we note that a large gap between $T_c$ and the
chemical freeze-out temperature $T$ with a high density hadronic phase
in between would make it very difficult to understand why all hadrons
freeze-out at the same temperature $T$.

Also included in Fig.~\ref{fig17} are calculations of freeze-out
curves for a hadron gas at constant energy density ($\varepsilon$=500
MeV/fm$^3$) and at constant total baryon density
($n_b$=0.12~fm$^{-3}$) \cite{pbm5}.  The LQCD phase boundary shown in
Fig.~\ref{fig17} is apparently departing at large $\mu_b$ from our
calculated curve of constant energy density, in contrast to the
(LQCD-based) results of ref. \cite{lqcd1} (see also \cite{taw1}).
However, the errors for the critical line from LQCD calculations
become rather significant for increasing $\mu_b$ \cite{lqcd1}.

The freeze-out points which are departing from the LQCD phase boundary
are reasonably well described by the curve of a hadron gas at constant
baryon density.  We note that this curve is very similar to the curve
corresponding to an average energy per average number of hadrons of
approximately 1~GeV \cite{1gev}.  

The underlying assumption of the thermal model used to extract the
($T$,$\mu_b$) values is (chemical) equilibrium at chemical
freeze-out. A natural question then is how is equilibrium achieved?
The answer obviously cannot come from within the framework of the
thermal model. Considerations about collisional rates and timescales
of the hadronic fireball expansion imply that at SPS and RHIC the
equilibrium cannot be established in a hadronic medium below the
critical temperature $T_c$ \cite{koch,greiner,huo,wetterich}. In a
recent paper \cite{wetterich} many-body collisions near $T_c$ were
investigated as a possible mechanism for the equilibration during
hadronization. There it is argued that, because of the rapid density
change near a phase transition, such multi-particle collisions provide
a natural explanation for the observation of chemical equilibration at
RHIC energies and it is suggested that $T=T_c$ to within an accuracy
of a few MeV.

One should note, however that, for e$^+e^-$ collisions, analyses of
hadron multiplicities \cite{aa08,becattini08} yield also a thermal-like pattern,
suggesting temperature values in the range of 160 MeV. Consequently,
alternative interpretations were put forward. These include
conjectures that chemical equilibrium is not reached by dynamical
equilibration among constituents but rather is a generic fingerprint
of hadronization \cite{stock,heinz}, or is a feature of the excited
QCD vacuum \cite{castorina}.  The situation for $e^+e^-$ collisions is,
however, quite different from that for nucleus-nucleus collisions: (i)
strangeness is not saturated, and (ii) additional, non-statistical
quantities such as the number of charm and beauty pairs in the system
play crucial roles in the description. We also remark that the
mechanism put forward in \cite{wetterich} is not equivalent to
'cooking' of hadrons in a hot and dense hadronic medium. Rather, the
phase transition plays the driving role. In that sense, the schematic
picture of \cite{stock,heinz} may not be so different from the
mechanism worked out in \cite{wetterich}.

These arguments enforce the conclusion that the hadro-chemical
freeze-out parameters probe experimentally in an unique manner the
critical line of the QCD phase transition between hadrons and QGP
\cite{js}. Our results imply that the phase boundary is reached
already for the SPS beam energy of 40 A GeV. Based on the LQCD results
of Fodor and Katz \cite{lqcd} shown in Fig.~\ref{fig17}, the
experimental freeze-out points at SPS are located in the vicinity of a
critical point found in this calculation. It was pointed out recently
\cite{op,op1} that the existence of a critical point for $\mu_b<$500
MeV requires a fine tuning of the (light) quarks masses within
5\%. However, it is important to recognize that serious open problems
of LQCD \cite{op,op1} need to be solved before one could address such
a delicate possibility and any connection between data and a possible
critical point remains tenuous at best.

In any case, the results discussed above imply that light (u,d,s)
quarks hadronize in a scheme which is goverend by thermal weights
determined by characteristic parameters $T$ and $\mu_b$ of the
fireball formed in ultra-relativistic nuclear collisions. Due to the
chemical equilibrium reached at this point, also all light quark abundances
must be thermal. 

We have noted above that chemical equilibrium is very likely not
reached for heavy quarks. Our conjecture which led to the statistical
hadronization model discussed next is that, while heavy quark
abundances are manifestly non-thermal but are determined mostly by
hard scattering processes, their subsequent hadronization is again
governed by the thermal weights at $T$ and $\mu_b$, i.e. at or very
near the phase boundary.

\section{The statistical hadronization model}

The statistical hadronization model (SHM)
\cite{pbm1,gor1,gra1,kos,aa1,aa2,aa3} assumes that 
the charm quarks are produced in primary hard collisions and that their total
number $N_c$ stays constant until hadronization. The constancy of $N_c$ has
been investigated in detail in \cite{aa2} where it is demonstrated that
variations of $N_c$ over the course of the evolution of the QGP are very
small.  Another important assumption is thermal equilibration in the QGP, at
least near the critical temperature, $T_c$. For a detailed discussion of this
issue see \cite{aa2}. Here we note only that recent data on flow and energy
loss of electrons from the decay of charmed hadrons lend support to the notion
that charm quarks apparently come close to thermal equilibrium in the QGP. For
a detailed discussion including possible elliptic flow of $J/\psi$ mesons see
\cite{bleicher08}.

Hadronization of all heavy quarks takes place near the phase boundary (see
above). The process of hadron formation for the bulk of heavy
quarks\footnote{We expect some fraction of nonthermalized partons, including
  heavy quarks, which would hadronize according to a, possibly modified,
  fragmentation function. This aspect is not covered by the approach discussed
  here.} is then not determined by fragmentation functions obtained from jet
studies \cite{pdg} but, in analogy to the situation for light quarks, is again
determined by weight functions obtained from a thermal ensemble with
temperature $T$, baryo-chemical potential $\mu_b$ and volume V. Our
hadronization prescription
is similar to the treatment of heavy flavor hadronization in thermal
descriptions of hadron production in $e^+e^-$ collisions
\cite{aa08,becattini08}, although in such systems there is no chemical
equilibration in the strangeness sector and several non-thermal quantities
need to be externally introduced into the calculations. Note, however, that charmonia
and bottomonia cannot be treated at all in this approach.

To introduce the strong departure from chemical equilibrium for charm
quarks we introduce a charm fugacity $g_c$ which regulates the number of charm
quarks in the thermal ensemble via the charm balance equation.

A key prediction of the statistical hadronization model is that the
$\chi_c/(J/\psi)$ and $\psi'/(J/\psi)$ ratios are entirely determined by the
hadronization temperature $T$. At $T$ = 164 MeV these ratios are much smaller
than what is observed in pp collisions \cite{pbm1}, implying that in such
systems charmonia are not produced via statistical hadronization. As we
discuss below, current data for $\psi'/(J/\psi)$ for already moderately
central Pb-Pb collisions at 
the top SPS energy are in good agreement with the statistical hadronization
model prediction. We note that the $\psi'/(J/\psi)$ ratio at RHIC and LHC energy
should be identical ($T$ does not change anymore) to the SPS result. The
observation of any serious discrepancy from the predicted value would be
difficult to explain within our model.

It is possible that hadrons containing bottom quarks and also bottomonia are
produced in a statistical hadronization picture just like charm quark. 
The consequences of such a hypothesis were indeed worked out in
\cite{gra2,aa2,bec_hq}. Whether bottom quarks thermalize in a hot and dense
fireball remains to be seen but upcoming experiments at the LHC will shed
light on this interesting issue.

\subsection{Procedures}

In the following we briefly outline the calculational steps in our model
\cite{pbm1,aa1}.  A much more detailed and technical discussion can be found
in \cite{aa2,aa3} to which we refer the reader for more information.  The
model has the following input parameters: i) the total number of charm quarks
$N_c$; ii) thermal characteristics of the fireball at chemical freeze-out:
temperature, $T$, baryo-chemical potential, $\mu_b$, and volume corresponding
to one unit of rapidity $V_{\Delta y=1}$.

The total number of charm quarks is ideally obtained by measuring the charm
production cross section in the nucleus-nucleus collision under consideration.
This way effects such as thermal charm production or shadowing are taken into
account explicitely. Unfortunately no such measurement exists today and in
practice we use the charm production cross section as measured or calculated
via perturbative QCD methods in pp collisions and extrapolate it to
nucleus-nucleus collisions assuming scaling with the number of hard
scatterings. Whenever available, estimates of shadowing are incorporated (see
below). All calculations of charmonia also include the effect of the nuclear
corona as discussed in \cite{aa2}. In the overlapping tails of the
nuclear density distributions, defined as regions with less than 10 \%
of the central density, we assume the physics is the same as in pp
collisions. 

The charm balance equation \cite{pbm1}, which has to include canonical
suppression factors \cite{gor1} whenever the number of charm quarks is not
much larger than 1, is used to determine a fugacity factor $g_c$ via:
\begin{equation}
N_{c\bar{c}}^{dir}=\frac{1}{2}g_c N_{oc}^{th}
\frac{I_1(g_cN_{oc}^{th})}{I_0(g_cN_{oc}^{th})} + g_c^2N_{c\bar c}^{th}.
\label{aa:eq1}
\end{equation}
Here $N_{c\bar{c}}^{dir}= N_c/2$ is the number of initially produced
$c\bar{c}$ pairs and $I_n$ are modified Bessel functions. In the fireball of
volume $V$ the total number of open ($N_{oc}^{th}=n_{oc}^{th}V$) and hidden
($N_{c\bar c}^{th}=n_{c\bar c}^{th}V$) charm hadrons is computed from their
grand-canonical densities $n_{oc}^{th}$ and $n_{c\bar c}^{th}$, respectively.
This charm balance equation is the implementation within our model of the
charm conservation constraint expressed in eq. (\ref{aa_eq0}) below.  The
densities of different particle species in the grand canonical ensemble are
calculated following the statistical model \cite{heppe,rhic,aa05}.  The
balance equation (\ref{aa:eq1}) defines the fugacity parameter $g_c$ that
accounts for deviations of heavy quark multiplicity from the value that is
expected in complete chemical equilibrium.  The yield of charmonia of type $j$
is obtained as: $N_j=g_c^2 N_j^{th}$, while the yield of open charm hadrons of
type $i$ is: $N_i=g_c N_i^{th}{I_1(g_cN_{oc}^{th})}/{I_0(g_cN_{oc}^{th})}$.

\subsection{On relevant time scales and medium effects}

In the original scenario of $J/\psi$ suppression via Debye screening
\cite{satz} it is assumed that the charmonia are rapidly formed in initial
hard collisions but are subsequently destroyed in the QGP. While it is clear
that the production of a (colored) charm quark pair takes place at time $t_{c
  \bar c} = 1/(2m_c) \leq 0.1$ fm, the formation time of charmonium involves
color neutralization and the build-up of its wave function. The relevant time
scale has been studied early on \cite{kar,bla} and is of order 1 fm. Similar
arguments also apply for the production time of hadrons with open charm and we
expect time scales comparable to those for charmonium.

We note that, at SPS energy where the 'melting scenario' was originally
studied, this time is in the same range as the plasma formation time. At SPS
and lower energies, charmonia can be formed in the pre-plasma phase and must
be destroyed in the plasma if suppression by QGP is to take place.

At the collider energies of RHIC and especially LHC the plasma formation time
is likely to be much shorter (comparable to $t_{c \bar c}$), implying that
charmonium or even color octet states do not exist before the plasma is
formed.  Furthermore, the number of charm quark pairs can exceed 10 per unit
rapidity. Initially, the 'collider' plasma will be hotter than T$_D$, the
temperature above which screening takes place, and no charmonia will be formed
at all in the QGP. It is our view that the charm quarks will be effectively
thermalized leading to an uncorrelated pool of c and $\bar c$ quarks.  Once
the plasma temperature falls below T$_D$ charmonia can be formed in principle,
but as is demonstrated by the studies performed in \cite{aa2}, their formation
rate is likely to be low.  Also we note that such a process would imply
hadronization of charm quarks at temperatures possibly much higher than the
critical temperature $T_c$. Considering the discussion about hadronization
above this is very unlikely. Similar arguments should apply also to the
hadronization of bottom quarks.

These observations lend further support to the notion, expressed explicitely
in the statistical hadronization model, that all charmonia are produced by
combination of charm and anticharm quarks at the phase boundary.  We would
like to emphasize that, in this scenario, the particular value of T$_D$ which
is much discussed in the recent literature \cite{satz2,moc}, is not very
important.  Even if $T_D$ turns out to be significantly larger than $T_c$, in
our view all charmonia are formed at the phase boundary.

Nevertheless, we note in passing that models combining the 'melting scenario'
with the statistical hadronization picture have been proposed \cite{gra1}.
Alternatively, charmonium formation by coalescence in the plasma
\cite{the1,the2,gra,yan} as well as within transport model approaches
\cite{zha,bra1} has been considered.

Another issue to be considered is the collision time
$t_{coll}=2R/\gamma_{cm}$, where $R$ is the radius of the (assumed equal mass)
nuclei and $\gamma_{cm}$ is the Lorentz $\gamma$ factor of each of the beams
in the center-of-mass system.  At SPS and lower energies, $\gamma_{cm} < 10$
and t$_{coll} > 1$ fm for a central Au-Au or Pb-Pb collision, so collision
time, plasma formation time, and charmonium formation time are all of the same
order. Furthermore, the maximum plasma temperature may not exceed T$_D$. In
this situation any possibly formed charmonia may be broken up only\footnote{We
  neglect break-up by hadronic co-movers, as unrealistic densities ($>
  1/fm^3$) are required to break-up charmonia and furthermore the comover
  mechanism leads to an incorrect rapidity dependence of charmonium
  production, as discussed in the next section.} by the high energy nucleons
still passing by from the collision and cold nuclear suppression needs to be
carefully considered, as discussed, e.g., in \cite{satz1,arleo}.  However, we
note in this context that the charm quarks resulting from such break-up
processes eventually have to hadronize, which might again lead to charmonium
production at the phase boundary if the charm quarks are kinetically
thermalized, as is assumed in the statistical hadronization model
\cite{pbm1,aa2}. Consequently, our calculations, in which both charmonium
formation before QGP production and cold nuclear suppression are neglected,
may somewhat underestimate the charmonium production yield at SPS energies and
below. In that sense the medium effects estimated below are upper limits for
energies close to threshold.

At collider energies there will be yet another separation of time scales. At
LHC energy, the momentum of a Pb nucleus is 2.76 TeV per nucleon, leading to
$p_{cm}$ =2.76 TeV and $\gamma_{cm} = 2940$, hence $t_{coll} < 5 \cdot
10^{-3}$ fm. Even ``wee'' partons with momentum fraction\footnote{We choose
  this value since it corresponds to a wee parton energy of approximately the
  binding energy of the J/$\psi$ meson.  Smaller x values are hence not
  relevant for the present considerations.}  $x_w = 2.5 \cdot 10^{-4}$ will
pass by within a time $t_w = 1/(x p_{cm}) < 0.3$ fm, and will not destroy any
charmonia since none exist at that time. We consequently expect that cold
nuclear absorption will decrease from SPS to RHIC energy and should be
negligible at LHC energy. First indications for this trend are visible in the
PHENIX data \cite{phe0}.

Given the various time scales it becomes clear from the above discussion that
the statistical hadronization model should become a quantitative tool to
describe charmonium and open charm production at collider energies without the
explicit need to take account of any charmonium or open charm hadron formation
before the QGP is developed and of cold nuclear absorption effects.  We note
in passing that the issue of shadowing or sa\-tu\-ra\-tion effects is of an entirely
different nature: within the framework of the statistical hadronization model
we need to know the rapidity density for open charm production in
nucleus-nucleus collisions. Using this quantity, which of course contains
shadowing or saturation effects, as input we can then provide cross sections
for the production of all open and hidden charm hadrons.

\section{Statistical hadronization model confronts charmonium data at SPS and
  RHIC energies}

Here we compare the results of statistical hadronization model calculations
with the most recent data from the SPS and RHIC. In the comparisons, the
$J/\psi$ yield is either normalized to the number of charm pairs or we use the
nuclear modification factor $R_{AA}^{J/\psi}$.

Here, $R_{AA}^{J/\psi}$ is defined as 
\be R_{AA}^{J/\psi}= \frac{\ud N_{J/\psi}^{AA}/\ud
    y}{N_{coll}\cdot\ud N_{J/\psi}^{pp}/\ud y} 
\ee 
and relates the charmonium yield in nucleus-nucleus collisions to that
expected for a superposition of independent nucleon-nucleon collisions.
In this expression, $\ud N_{J/\psi}/\ud y$ is the rapidity density of the
$J/\psi$ yield integrated over transverse momentum and $N_{coll}$ is the
number of binary collisions for a given centrality class. The
hadronization volume always covers one unit of rapidity for the
calculations shown in this article. 

\begin{figure}[ht]
\hspace{-.7cm}
\begin{tabular}{lr}
\begin{minipage}{.49\textwidth}
\vspace{-.6cm}
\centering\includegraphics[width=1.2\textwidth]{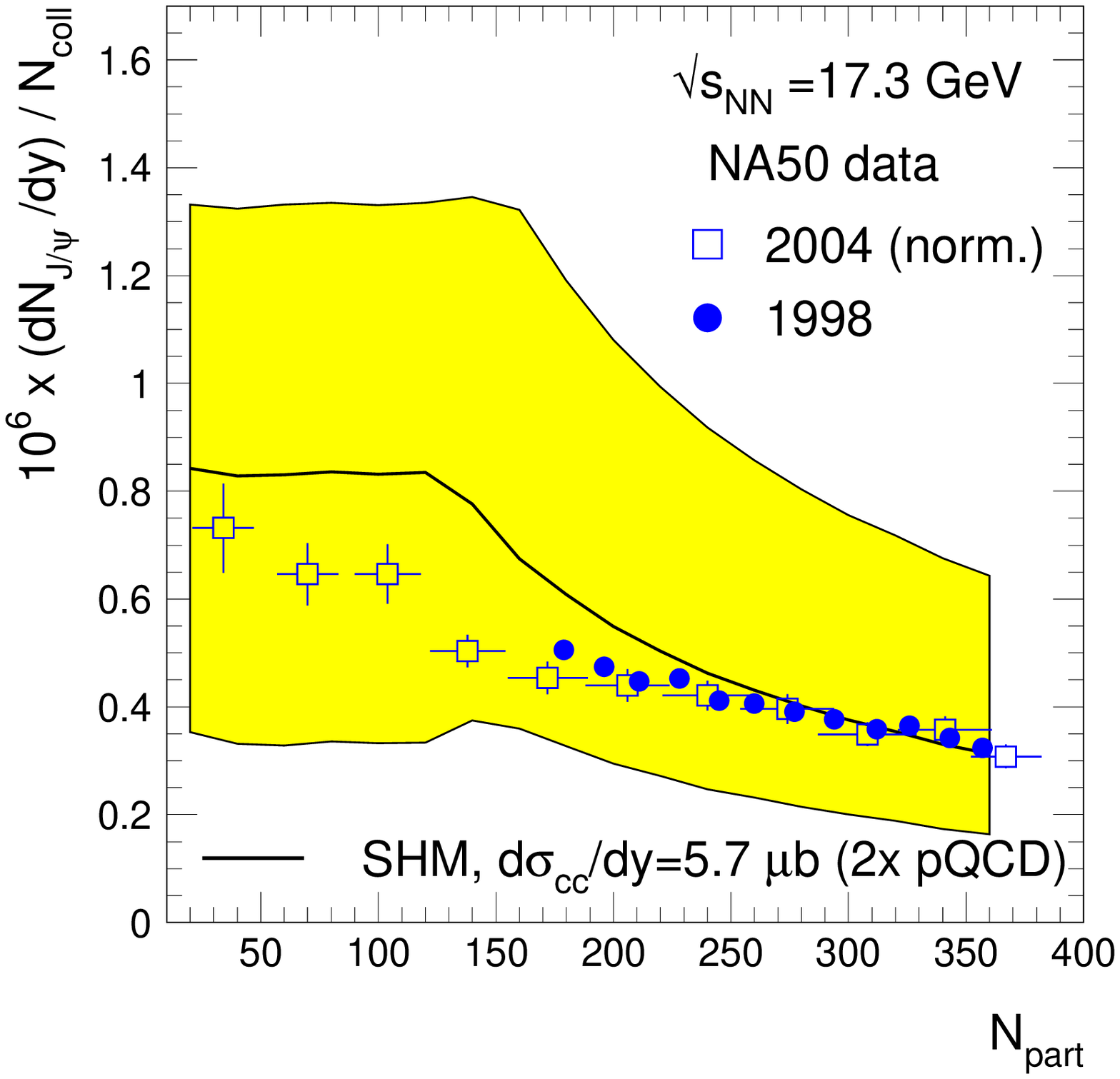}
\end{minipage} \begin{minipage}{.49\textwidth}
\vspace{-.6cm}
\centering\includegraphics[width=1.2\textwidth]{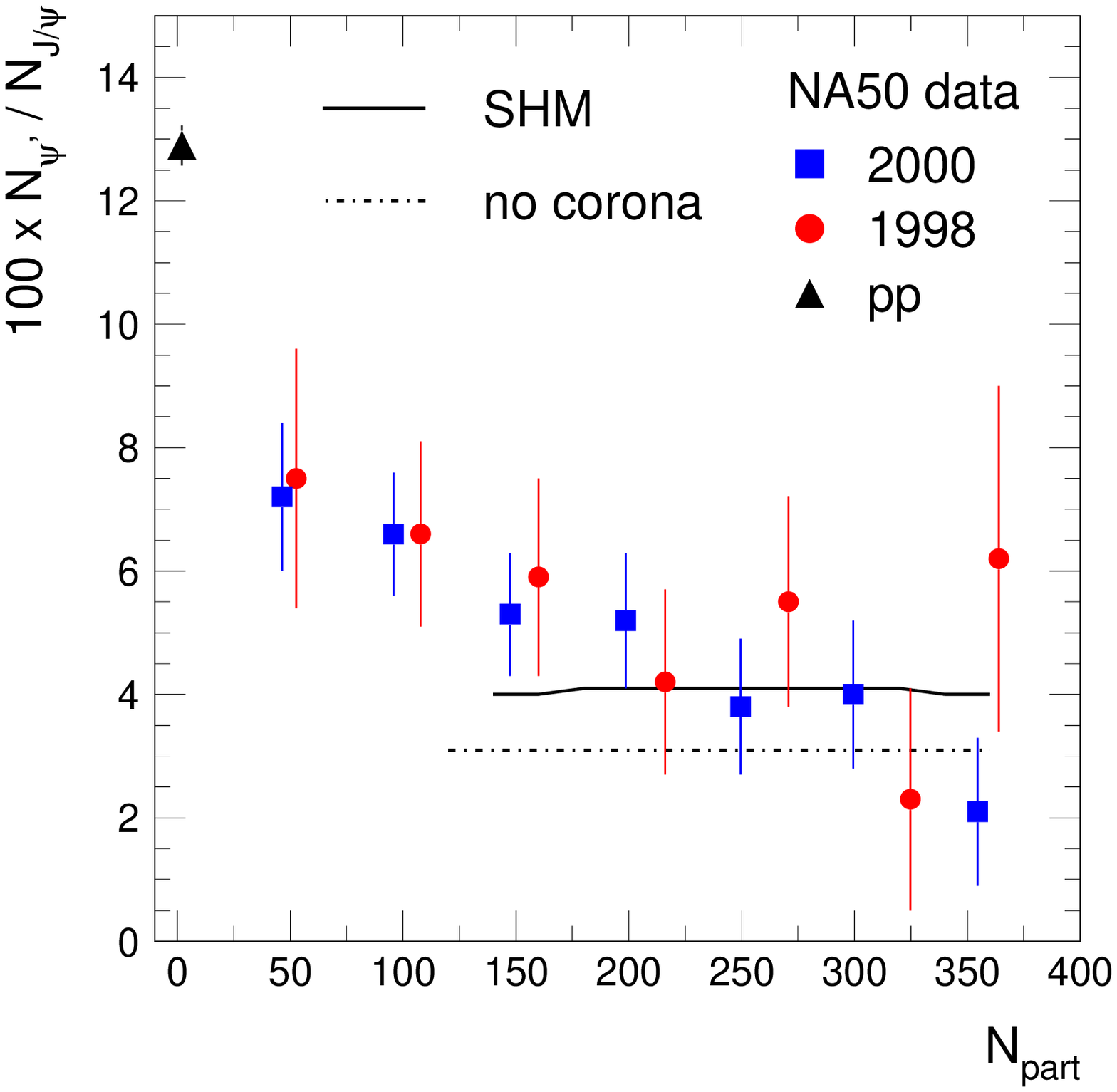}
\end{minipage}
\end{tabular}
\caption{Centrality dependence of $J/\psi$ yield per number of
  collisions (left panel) and of the ratio of yields of ${\psi'}$ and
  ${J/\psi}$ (right panel) at top SPS energy. The shaded band in the
  left panel includes the assumed range in the charm production cross
  section (a factor of 2 up and down) and the error assumed for the
  $J/\psi$ cross section in pp collisions.  The experimental data on
  $J/\psi$ yield are from ref. \cite{gos} (see \cite{pbm1}) and
  ref. \cite{na50a} (2004 NA50 data) and on the ${\psi'}/{J/\psi}$
  ratio from ref. \cite{na50b,na50c}. Figure taken from \cite{aa2}.}
\label{aa_fig3a}
\end{figure}

\subsection{SPS data}

The comparison of our model calculations with the data measured at the
SPS by the NA50 experiment is shown in Fig.~\ref{aa_fig3a}.

The experimental data on $J/\psi$ production per binary collision
(left panel) are well described by invoking a moderate enhancement of
the charm production cross section of a factor of 2 compared to pQCD
calculations \cite{rv1}.  New measurements by the NA60 experiment
\cite{na60a} indicate that the enhancement in the dimuon yield below
the $J/\psi$ mass, earlier observed by NA50 \cite{na50en}, is of
prompt origin and, as such, cannot be interpreted as an enhancement of
the charm production cross section, although an experimental result on
the charm cross section at this energy is currently not available.  We
also note that a factor of 2 enhancement is within uncertainties of
the pQCD calculations.

A comparison between calculations and data \cite{na50b,na50c} is shown in the
right panel of Fig.~\ref{aa_fig3a} for the centrality dependence of the yield
of ${\psi'}$ relative to $J/\psi$. Taking account of the corona effect leads
to a 25\% increase of the $\psi'/(J/\psi)$ yield ratio over the value
using Boltzmann factors only. Already for moderately central collisions the
data approach the thermal value of the statistical hadronization model. The
$\psi'/(J/\psi)$ yield as measured in pp collisions is shown as a triangle.

\subsection{RHIC data}

\begin{figure}[ht]
\hspace{-.7cm}
\begin{tabular}{lr}
\begin{minipage}{.49\textwidth}
\vspace{-1cm}
\centering\includegraphics[width=1.2\textwidth]{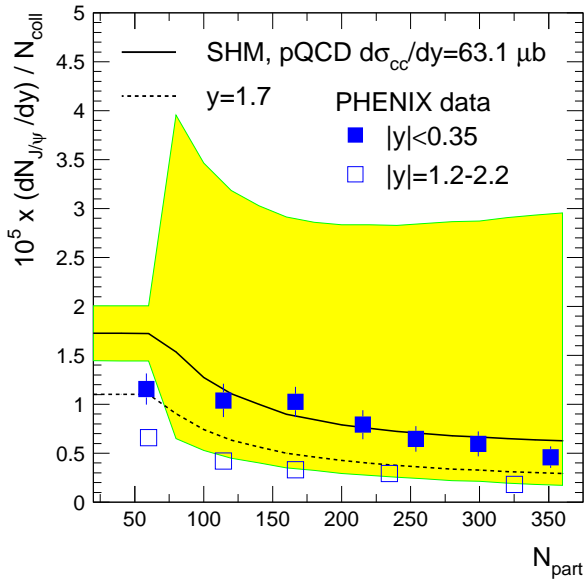}
\end{minipage}  & \begin{minipage}{.49\textwidth}
\vspace{-1cm}
\caption{Centrality dependence of the $J/\psi$ yield at midrapidity
  and in a more forward/backward rapidity bin per number of collisions
  at RHIC \cite{phe1}. The experimental data are compared to model
  calculations, using the charm production cross section as calculated
  in pQCD \cite{cac}.  The line is the calculation for the central
  value of the charm production cross section, the band corresponds to
  its systematic uncertainty. The dashed line is the model prediction
  for rapidity $y$=1.7. Figure taken from \cite{aa2}.}
\label{aa_fig3}
\end{minipage}
\end{tabular}
\end{figure}

The centrality dependence of $R_{AA}^{J/\psi}$ at midrapidity as well
as at more forward/backward rapidity is shown in
Fig.~\ref{aa_fig3}.  The model reproduces very well the decreasing
trend versus centrality seen in the RHIC data \cite{phe1} and also the
relative ratio of the two rapidity intervals. Note that, in our model,
the centrality dependence of the nuclear modification factor arises
entirely as a consequence of the still rather moderate rapidity
density of initially produced charm quark pairs at RHIC. 

 The statistical hadronization model calculations reproduce naturally
 the unexpected result that the observed $J/\psi$ suppression at the
 SPS and at RHIC are rather similar, even though the energy density of
 fireballs produced in central Au-Au collisions at RHIC (about 5.5
 GeV/fm$^3$) significantly exceeds that produced in Pb-Pb collisions
 at SPS energy (about 3 GeV/fm$^3$. Here, we think the RHIC value is
 conservative, since for both energies a proper time of 1 fm was used
 to estimate the energy density. For a more detailed discussion see
 \cite{js}.

We note that the calculations shown in Fig.~\ref{aa_fig3} use the
charm production cross section of \cite{cac}. This is about a factor
of 2 smaller than the charm cross sections derived by the PHENIX
collaboration from single electron measurements \cite{cc2} although that result
is also consistent with the pQCD value of \cite{cac} due to the large error
bands
\footnote{Charm cross sections given by the STAR collaboration are larger
  by about another factor of 2 \cite{star_charm}. Using these in the
  statistical hadronization model leads to a significant overestimate of the
  $J/\psi$ data.}. Shadowing effects may further reduce the total charm yield
in nuclear collisions, see below.  

\begin{figure}[htb]
\vspace{-.5cm}
\centering\includegraphics[width=.84\textwidth]{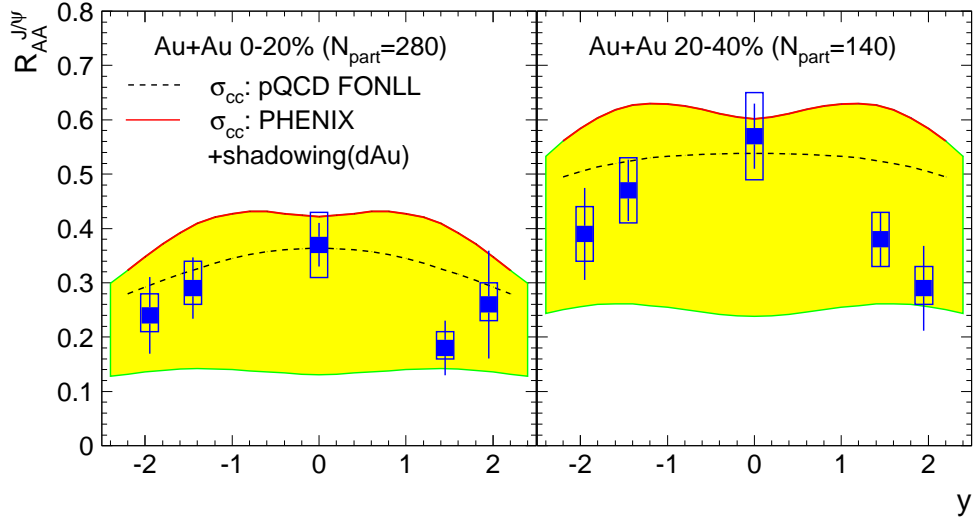}
\vspace{-.5cm}
  \caption{Rapidity dependence of $R_{AA}^{J/\psi}$ for two centrality
    classes. The data from the PHENIX experiment \cite{phe1} (symbols
    with errors) are compared to calculations (dashed and solid red
    lines, see text). The shaded area corresponds to calculations
    covering down to the lower limit (green line) of the charm cross
    section as measured by PHENIX \cite{cc2}, and using our shadowing
    scenario. Figure taken from \cite{aaqm08}.}
\label{aa_fig1}
\end{figure}

Equally striking are the results on the rapidity dependence of the
nuclear modification factor $R_{AA}^{J/\psi}$ observed by the PHENIX
collaboration \cite{phe1}. They are already visible in
Fig.~\ref{aa_fig3} and are explicity presented in
Fig.~\ref{aa_fig1}. Contrary to expectations based on the assumption
that $J/\psi$ suppression should scale with energy density (see, e.g.,
Fig. 9 in \cite{na50a}), the experimentally observed suppression is
smallest at midrapidity, where the energy density is largest, but
increases towards forward and backward rapidities.

The observed rapidity dependence is equally at odds with predictions
for charmonium suppression within the comover approach
\cite{gavin,capella,urqmd_jpsi} since the largest comover density is
at midrapidity.

A maximum in the nuclear modification factor a midrapidity emerges
naturally within the framework of the statistical hadronization
model. This is due to the enhanced generation of charmonium around
mid-rapidity, determined by the rapidity dependence of the charm
production cross section which has a maximum at midrapidity. Our
calculations on the rapidity dependence of the nuclear modification
factor $R_{AA}^{J/\psi}$ \cite{aaqm08} are compared to the
experimental data in Fig.~\ref{aa_fig1}.  The dashed line in this
figure is, as above, obtained using the pQCD charm production cross
section \cite{cac}. The upper (red) solid line is based on the
currently available experimental input into charm production at RHIC
energy. It is based on the charm cross section derived by PHENIX
in pp collisions \cite{cc2} modified to take into account possible
shadowing effects in Au-Au collisions. To this end we assume that the
deviation from unity of the nuclear suppression factor in d-Au
measurements \cite{phe2}, $R_{dAu}^{J/\psi}$, is due to shadowing and
compute the shadowing factor in Au-Au collisions as 
\be
S_{Au-Au}(y)=R_{dAu}^{J/\psi}(y) R_{dAu}^{J/\psi}(-y).  
\ee 
Using this as input into the statistical hadronization model
calculation leads to a good description of the observed suppression
and its rapidity dependence, as anticipated.  We take this result as
first evidence for the statistical generation of J/$\psi$ at chemical
freeze-out, as expected in the statistical hadronization model.

Nevertheless, there are alternate explanations of the RHIC data on $J/\psi$
production, as summarized recently \cite{raphael_gr}. We note in particular
efforts incorporating very large gluon saturation effects as expected in the
framework of the color-glass model \cite{glass1,glass2} and gluon shadowing
effects \cite{bravina_ca}.  Also the possible
influence on charmonium production of a strongly coupled quark-gluon plasma
has been investigated \cite{young_sh}.  Much improved data on charmonium and
open charm production will be needed to sort out these different approaches at
RHIC energy. In particular, our approach of statistical hadronization for
charmonium production is very sensitive to the charm cross section as the only
input parameter due to the quadratic dependence. A measurement accuracy of 10
\% would be highly desirable and significant experimental progress it expected
to come with the upgrades of PHENIX and STAR with vertex detectors.

\section{Effects of medium modifications on charmed hadron and charmonium production}

In the following, we explore charmonium and charmed hadron production
and its sensitivity to medium modifications from the low energy domain
near threshold ($\sqrt{s_{NN}} \approx 6$ GeV) to top SPS
and RHIC energies. The lower part of this energy range can be
investigated in the CBM experiment \cite{cbm1} at the future FAIR
facility. One of the motivations for such studies was the
expectation\cite{cbm1,tol} to provide, by a measurement of D-meson
production near threshold, information on their possible modification
near the phase boundary.  Here we demonstrate that, because of the
relevant mass and time scales involved, medium effects on D-meson
production are likely to be very small. Furthermore, because of the
dominance of associated charm production at low energies, it turns out
to be important to measure in addition to D-meson production also the
yield of charmed baryons to get a good measure of the total charm
production cross section. Most of the arguments presented below are
not a strong function of energy and apply to higher energies as
well. Possible effects due to the charm threshold at energies below
$\sqrt{s_{NN}} \approx 8$ GeV are neglected. Our review is based on the recent
work of \cite{charm_le}.

\begin{figure}[ht]
\vspace{-1cm}
\centering\includegraphics[width=.83\textwidth,height=.75\textwidth]{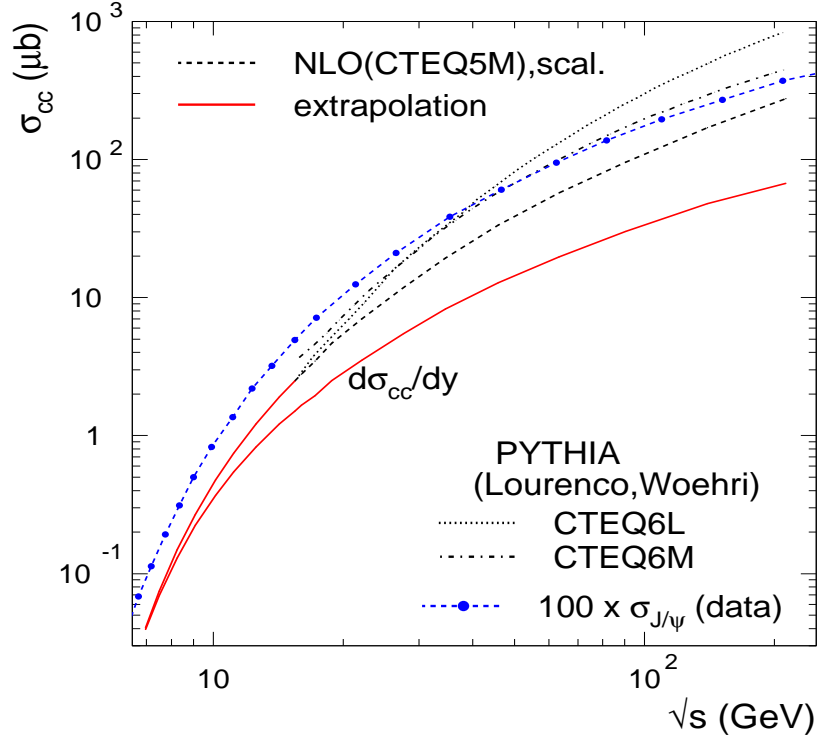}
\caption{Energy dependence of the charm production cross section in pp
  collisions. The NLO pQCD values of Vogt \cite{rv1} are compared to
  calculations using PYTHIA, taken from ref. \cite{lou}.  Our
  extrapolations for low energies are shown as solid lines, for total
  and midrapidity ($\ud\sigma_{c\bar{c}}/\ud y$) cross sections.  The
  line marked with (blue) dots indicates a parameterization of the
  measured energy dependence of the $J/\psi$ production cross section
  \cite{herab}.}
\label{aa_fig0}
\end{figure}

For an investigation of charmonium production below top SPS energy we
need an estimate of the energy dependence of the total number of
produced charm quarks $N_c$.  As no information on the charm
production cross section is available for energies below
$\sqrt{s_{NN}}$=15 GeV, we have to rely on extrapolation. As basis for
this extrapolation we take the energy dependence of the total charm
production cross section calculated in ref. \cite{rv1}, as shown in
Fig.~\ref{aa_fig0} \footnote{We have rescaled these calculations to
  more recent values calculated at $\sqrt{s_{NN}}$=200 GeV in FONNL in
  ref.  \cite{cac}.}. The extrapolated curves for charm production
cross section are shown as solid lines in Fig.~\ref{aa_fig0}.
Also shown for comparison are calculations with PYTHIA \cite{lou}.

The total charm production cross section has an energy dependence
similar to that measured for $J/\psi$ production by the HERA-B
collaboration \cite{herab}, also shown in Fig.~\ref{aa_fig0}. The
extrapolation procedure for the low-energy part of the cross section
obviously implies significant uncertainties. We emphasize, however,
that the most robust predictions of our model, i.e. the yields of
charmed hadrons and charmonia relative to the initially produced $c
\bar c$ pair yield are not influenced by the details of this
extrapolation.

\subsection{Charm conservation equation}

Charm conservation is essential for the discussion of possible
in-medium changes of charmed hadrons and their effect on the
production cross sections. We start the discussion by recalling that
\be 
\sigma_{c \bar c} = \frac{1}{2} ( \sigma_D + \sigma_{\Lambda_c}
+\sigma_{\Xi_c} + ...) + ( \sigma_{\eta_c} + \sigma_{J/\psi} +
\sigma_{\chi_c} + ...)
\label{aa_eq0}  
\ee 
because of charm conservation. As shown in \cite{aa2}, annihilation
of charm quarks can be neglected.  In the above equation,
$\sigma_D$ is the total cross section for the production of any
D-meson. As discussed above, the cross section $\sigma_{c \bar c}$ is
governed by the mass of the charm quark $m_c \approx 1.3$ GeV
\cite{pdg}. Consequently, perturbative QCD procedures can be applied
even at low energy.  Since charm production is a 'hard process' we
expect no medium effects on the left-hand side of eq.~\ref{aa_eq0}
and, hence, there are no medium effects on the hadronic side of
eq.~\ref{aa_eq0} either.  Such a separation of scales is not possible
for strangeness production, and the situation there is not easily
comparable. This also implies that there is no clear relation between
near-threshold kaon production at SIS18 energy and near threshold
D-meson production at FAIR.

The much later formed D-mesons, or other charmed hadrons, may well
change their mass in the hot medium. The results of various studies on
in-medium modification of charmed hadrons masses
\cite{tol,tsu,sib1,sib,hay,cas,fri,lutz} are sometimes contradictory
and we refer to \cite{charm_le} for a detailed assessment.

Whatever the medium effects may be, they can, because of the charm
conservation equation above, only lead to a redistribution of charm
quarks among the various hadronic final states.  In particular, if
the masses of all D-mesons are lowered by the same amount at the phase
boundary, this effect would practically not be visible in the D-meson
yield.  Although the charm conservation equation above is strictly
correct only for the total cross section we expect, within the
framework of the statistical hadronization model, also little
influence due to medium effects on distributions in rapidity.  This is
due to the fact that the crucial input into our model is $\ud
N_{c\bar{c}}^{AuAu}/\ud y$ and there is no substantial D-meson
rescattering after formation at the phase boundary.  Modification of
D-meson masses at the phase boundary will, however, influence the
production rates for charmonia: after lowering of their masses the
D-mesons will ``eat away'' the charm quarks which might form charmonia but since the
D-mesons are much more abundant, their own yield will hardly change.

Another caveat is that medium modifications are only visible within
our approach if the charmed hadrons have changed masses at chemical
freeze-out.  We note that excellent fits of the common (non-charmed)
hadrons to predictions of the thermal model have been obtained using
vacuum masses \cite{heppe,rhic,aa05}. In order to see the maximum
possible effect of in-medium modification, we neglect this caveat in
the following and compute thermal weights for charmed hadrons with
masses different from their vacuum values.
   
Much of the above argument about medium effects is essentially
model-independent and applies equally well at all energies.  Here we
will consider various types of scenarios for medium modifications and
study their effect within the statistical hadronization framework in
the energy range from charm threshold to collider energies.

\begin{figure}[hbt]
\centering\includegraphics[width=.66\textwidth]{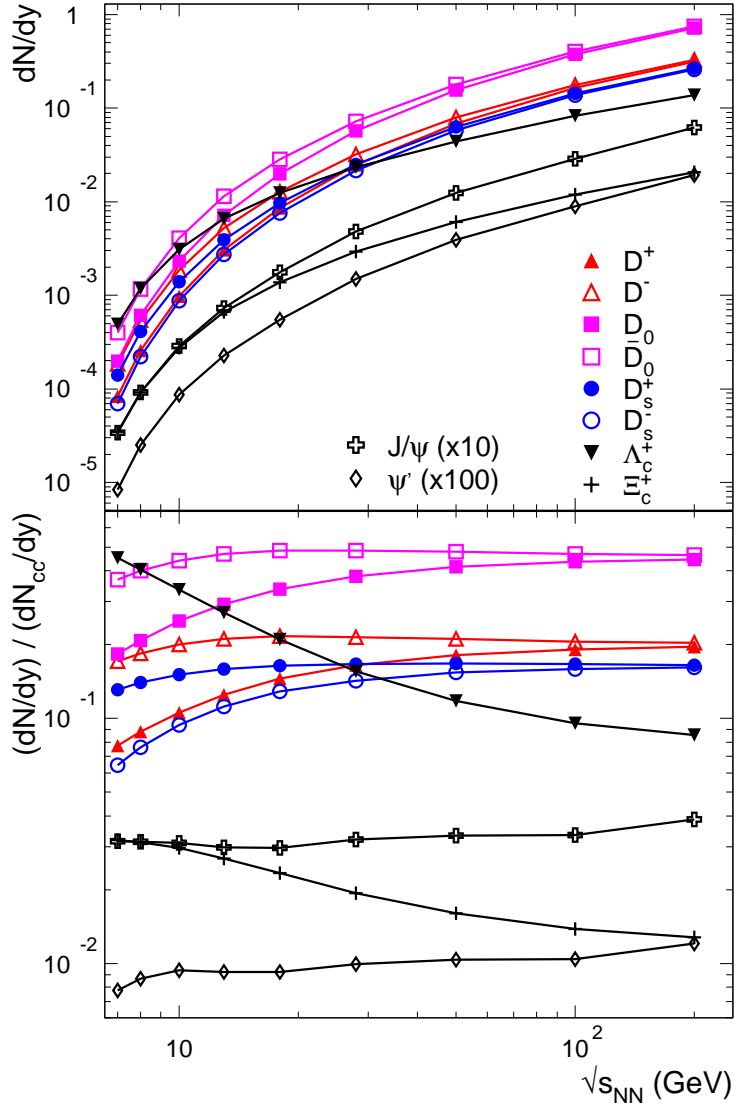}
\caption{Energy dependence of charmed hadron production at midrapidity
  computed with the statistical hadronization model. Upper panel:
  rapidity densities, lower panel: rapidity densities relative to the
  number of $c\bar{c}$ pairs. Note the scale factors of 10 and 100 for
  $J/\psi$ and $\psi'$ mesons, respectively. Figure taken from \cite{charm_le}.}
\label{le_fig1}
\end{figure} 

\subsection{Results}

As a first result we present, in Fig.~\ref{le_fig1}, the situation for
vacuum masses.  The upper panel shows our predictions for the energy
dependence of rapidity densities for various charmed hadrons, while in
the bottom we have normalized to the number of initially produced
pairs $c\bar{c}$. Beyond the generally decreasing trend for all yields with
decreasing beam energy
one notices first a striking behavior of the production of $\Lambda_c$
baryons: their yield exhibits a much weaker energy dependence than
observed for other charmed hadrons. This is a result of the increase
in baryo-chemical potential with decreasing energy.  A similar
behavior is seen for the $\Xi_c^+$ baryon.  In detail, the production
yields of D-mesons depend also on their quark content.

\begin{figure}[hbt]
\centering\includegraphics[width=.77\textwidth]{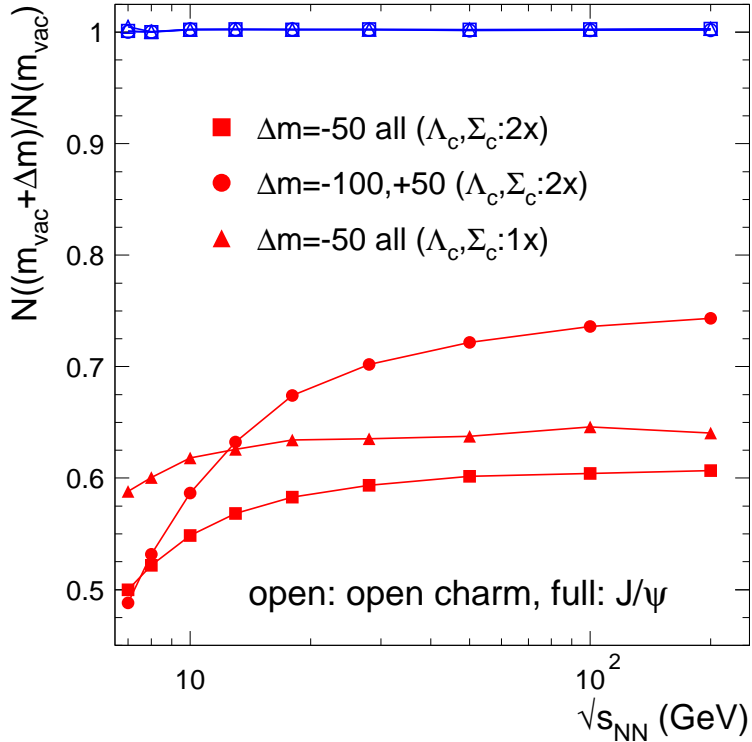}
\caption{Energy dependence of the relative change in the yield of open
  charm hadrons and of $J/\psi$ meson considering different scenarios
  for in-medium mass modifications. The squares correspond to a
  lowering in mass by 50 MeV for charmed hadrons except for
  $\Lambda_c$ and $\Sigma_c$ where double the shift is assumed. The
circles assume a shift downward for hadrons with charm quarks and
upwards for hadrons with anticharm quarks and again double the change
for $\Lambda_c$ and $\Sigma_c$. The triangles correspond to a
  downward shift in mass of 50 MeV for all charmed hadrons. Figure taken from \cite{charm_le}.}
\label{le_fig2}
\end{figure}

We note that the model prediction of yields relative to $c\bar{c}$
pairs is a robust result, as it is to first order independent of the
poorly known absolute charm production cross section. The results
shown in Fig.~\ref{le_fig1} and Fig.~\ref{le_fig2} could be extrapolated up the LHC energy
without visible change due to the expected unchanged
temperature\cite{aa4}.

The results expected if one introduces mass changes are presented in
Fig.~\ref{le_fig2} for three different scenarios. In this figure we
demonstrate that the total open charm yield (sum over all charmed
hadrons) exhibits essentially no change if one considers mass shifts,
while the effect is significant on charmonia. This is to be expected
from eq. (\ref{aa:eq1}): as D-meson and $\Lambda_c$-baryon masses are
reduced, the charm fugacity g$_c$ is changed accordingly to conserve
charm.  Since the D-meson and $\Lambda_c$-baryon yields vary linearly
with g$_c$ we expect little change, while yields of charmonia vary
more strongly, since their yields are proportional to $g_c^2$.

Note that the reduction of the $J/\psi$ yield in our model is quite different
from that assumed in \cite{gra,hay,sib,fri}, where a reduction in
D-meson masses leads to (in-medium) decay of $\psi'$ and $\chi_c$ into 
$D\bar{D}$.

\section{Charmonium production at LHC energy} 

Here we present our predictions \cite{aa2,aa3,aaqm08,aa4} for charmonium
production at LHC energy.  At LHC energy the charm production cross section
(including shadowing in PbPb collisions \cite{rv1}) is expected to be about an
order of magnitude larger than the value measured at RHIC .  With 
$\frac{\ud N_{c \bar c}^{PbPb}} {\ud y} \geq 10$ in central collisions canonical effects
  in the thermal charm calculation are expected to be small (see section 3.1)
  implying from eq.~(\ref{aa:eq1}) that the charm fugacity $g_c \propto N_c$ and,
  hence, charmonium production will scale quadratically with $N_c$, at least
  for central collisions.  As a result, the nuclear modification factor for
  charmonium production should exhibit a trend as function of
  centrality which is qualitatively different from results observed at RHIC: 
  Depending on the actual charm production cross section $R_{AA}^{J/\psi}$
  should be flat or increase with increasing number of 
  participants $N_{part}$, i.e. as the collisions get more central. For an
  actual  charm production cross section above 640 $\mu$b
  $R_{AA}^{J/\psi}$ should even exceed  unity for central  
  collisions.

\begin{figure}[htb]
\begin{tabular}{cc}
\begin{minipage}{.49\textwidth}
\centering\includegraphics[width=.94\textwidth]{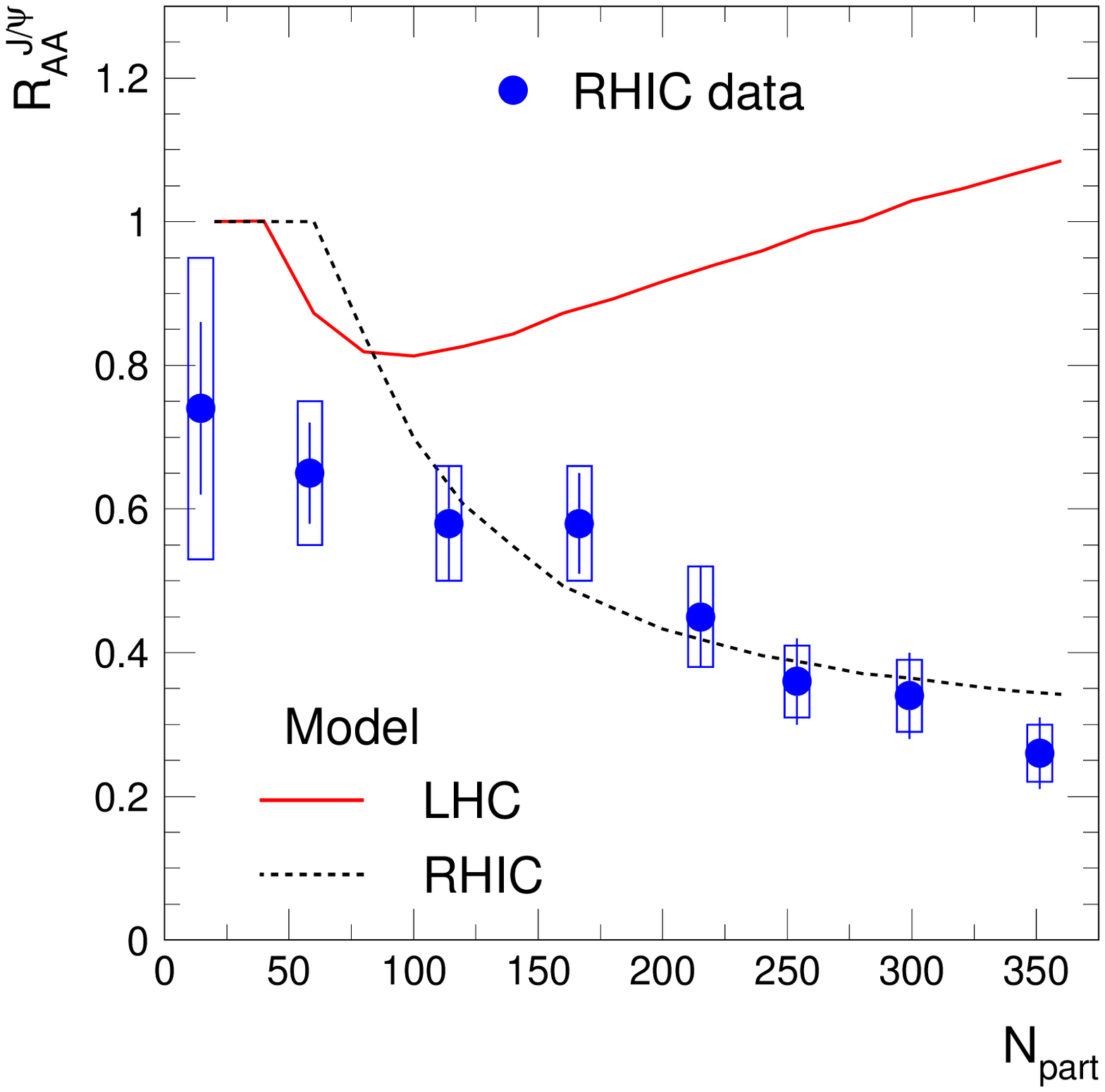}
\end{minipage}  & \begin{minipage}{.49\textwidth}
\centering\includegraphics[width=.94\textwidth]{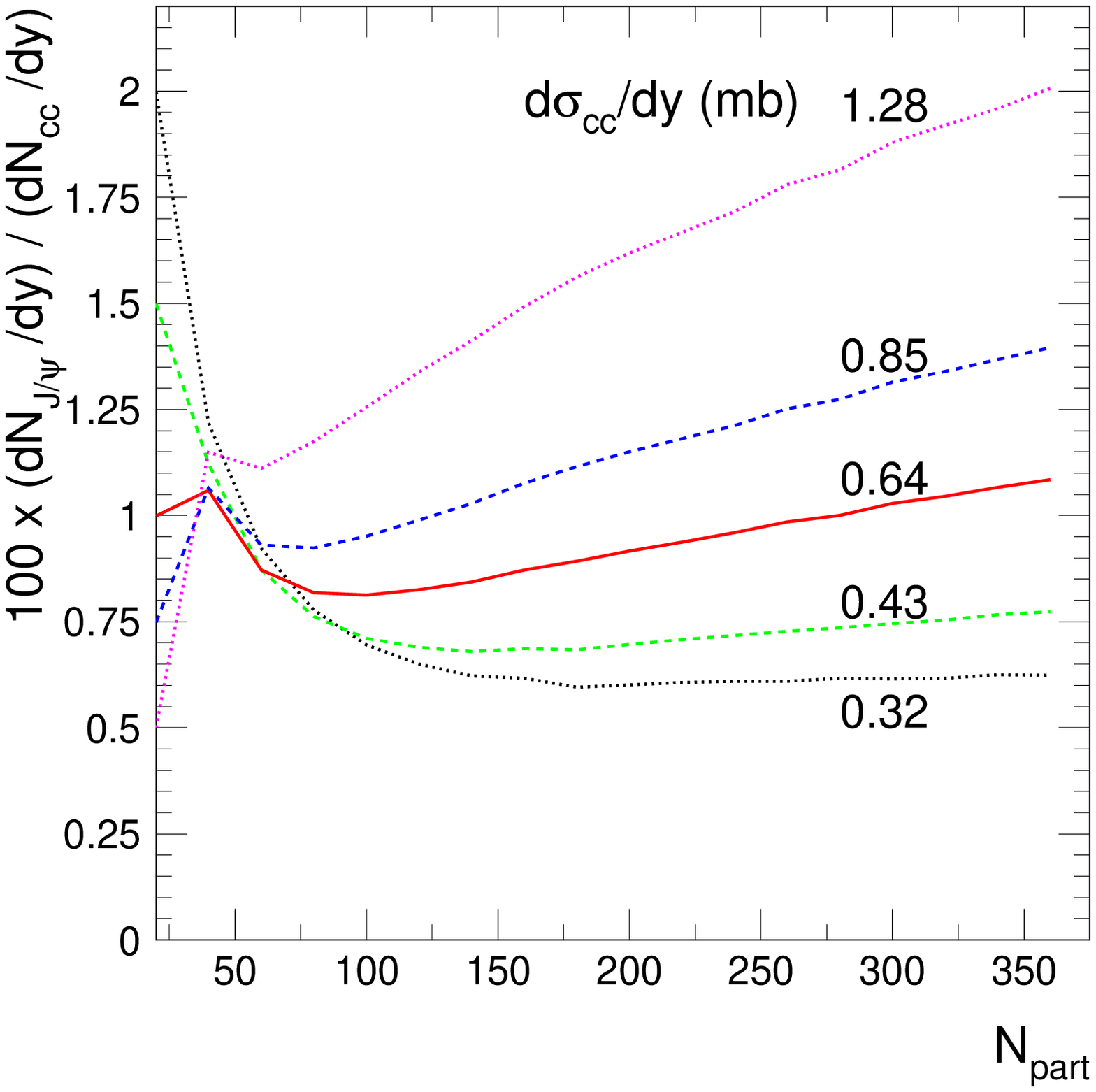}
\end{minipage}\end{tabular}
\vspace{-.5cm}
\caption{Centrality dependence of $R_{AA}^{J/\psi}$ for RHIC and LHC energies
(left panel) and of the $J/\psi$ rapidity density  at midrapidity at LHC relative to the
number  
of initially produced $c\bar{c}$ pairs (right panel, curves labelled by the
$\ud\sigma_{c\bar{c}}/\ud y$). Figure taken from \cite{aaqm08}.}
\label{aa_fig2}
\end{figure}

 This is shown in Fig.\ref{aa_fig2}. In the
 left panel the centrality dependence of $R_{AA}^{J/\psi}$ at LHC
 energy is calculated using charm cross sections predicted in
 \cite{rv1}, while the comparison to charmonium data from RHIC is done
 using the cross section computed for pp collisions by \cite{cac} and
 scaled with the number of hard collisions (see section 4.2). The dramatic difference
 between RHIC results and LHC predictions is obvious. We note,
 however, that the degree of charmonium suppression or enhancement in
 PbPb collisions at LHC energy depends strongly on the open charm
 cross section. This is demonstrated in the right panel of
 Fig.~\ref{aa_fig2}, where we show the centrality dependence of
 $J/\psi$ production for a reasonable range in the number of initially
 produced $c\bar{c}$ pairs\footnote{Note that $R_{AA}^{J/\psi}
 N^{J/\psi}_{pp}/N^{c \bar c}_{pp} = N^{J/\psi}_{AA}/N^{c \bar c}_{AA}$.}. 
Over this range of charm cross section the
 yield of charmonia could vary by more than  a factor of 10 and it is
 clear from this figure that a reasonably precise measurement of the
 open charm cross section in PbPb collisions is required to settle
 quantitatively the issue of statistical hadronization. The
 observation of an increase with $N_{part}$ of the charmonium yield
 will, however, be already a strong indication that charmonium
 production through combination of charm and anticharm quarks at the
 phase boundary is observed. This would be a spectacular fingerprint
 of deconfined and thermalized charm quarks in the quark-gluon plasma.

\section{Summary and outlook}

We have reviewed the physical basis for a description of charmonium production
in ultra-relativistic nuclear collisions within the framework of the
statistical hadronization model. Starting from an analysis of thermal
production of hadrons composed of light (u,d,s) quarks we have presented the
by now convincing evidence that these hadrons are formed (nearly)
simultaneously at the QCD phase boundary by a process called ``chemical
freeze-out'' with hadronic abundances quantitatively described by an
equilibrated statistical ensemble. In other words, hadronization of light
quarks takes place during the phase transition from the quark-gluon plasma to
hadronic matter.

Heavy quarks, such as charm and bottom quarks, are not produced  in the
thermal fireballs formed in nucleus-nucleus collisions at high energy, but
result dominantly from initial, hard collisions. Owing to
their large mass which exceeds the QCD phase transition temperature by more
than a factor of 5 for charm quarks and 20 for bottom quarks, chemical
equilibrium is never attained for heavy quarks. However, there is by now
mounting evidence from studies of charm quark flow and energy loss, that
thermal equilibrium is reached in the charm quark sector. This also implies
that charm quarks can travel significant distances to combine with
uncorrelated anti-charm quarks. Experimental confirmation of predictions from
the statistical hadronization model would then also imply direct information
on the degree of deconfinement reached in the fireball. 

An interesting and
completely open question is whether thermal equilibrium is also reached for
bottom quarks. Measurements at the LHC of possible flow and energy loss of
identified bottom quarks will be crucial to shed more light on this issue.

We have described the development of the
statistical hadronization model for heavy quarks and its application to
charmonium production. In this model, all charm quarks are produced
non-thermally but, in close analogy to the situation in the light quark
sector, hadronize at the QCD phase boundary. Open and hidden charm hadrons are
then formed with relative abundances again determined by statistical weight
factors which can be computed again using a statistical ensemble as for the
light quarks. We note that hadronization of heavy quarks at temperatures
higher than the QCD critical temperature is not compatible with this view.
Analyses of data on charmonium production at SPS and RHIC energies are in good
agreement with predictions from the statistical hadronization model. Further,
more detailed tests at RHIC energy and, in particular, analysis of data at LHC
energy  are essential to identify uniquely the underlying physics.

Using the charm sector to search for possible medium effects, i.e. mass
modifications of charmed hadrons in the hot fireball, is severely constrained
by charm conservation. This implies, e.g., that cross sections for D-meson
production are very insensitive to the actual mass of D-mesons since medium
modifications will modify the distribution of charm quarks among the different
hadron species while the initial charm production process is governed by the
mass of the charm quark.

A crucial aspect of the statistical hadronization model and, in our opinion,
of all models on charmonium production in ultra-relativistic nuclear
collisions is to take account of the time sequence of events from the
production of charm quarks to the formation of charmonia. We have discussed
this issue in detail.
In particular, the formation time of open and hidden charm hadrons is
important and, especially at collider energies, likely longer than the
collision time or QGP formation time. This provides further support for the
notion of late charmonium production at the phase boundary.

A key feature of the statistical hadronization approach is the scaling of
charmonium with the square of the number of charm quark pairs in the fireball.
We have argued that first traces of this scaling are already visible in the
RHIC data on charmonium production. Data from the LHC will provide a crucial
test for this approach.

\section{Acknowledgments}
The authors would like to recognize the continued and very fruitful
collaboration with A. Andronic and K. Redlich, with whom many of the ideas and
results discussed in this review have been worked out. pbm acknowledges the
support of the German Helmholtz Society through its alliance program.

\end{document}